\documentstyle[epsf,psfig]{mn}
\begin{document}
\title[Formation times and masses]
      {Formation times and masses of dark matter haloes}
\author[R. K. Sheth \& G. Tormen]{Ravi K. Sheth$^1$ \& Giuseppe Tormen$^2$\\
$^1$ Departement of Physics \& Astronomy, University of Pittsburgh, 
     3941 O'Hara Street, PA 15260, USA\\
$^2$ Dipartimento di Astronomia, Vicolo dell'Osservatorio 2, 
     35122 Padova, Italy \\
\smallskip
Email: rks12@pitt.edu, tormen@pd.astro.it
}
\date{Submitted 6 January 2004; in original form 15 October 2003}

\maketitle

\begin{abstract}
The most commonly used definition of halo formation is the time when 
a halo's most massive progenitor first contains at least half the 
final mass of its parent.  
Reasonably accurate formulae for the distribution of formation times 
of haloes of fixed mass have been available for some time.  
We use numerical simulations of hierarchical gravitational clustering 
to test the accuracy of formulae for the mass at formation.  
We also derive and test a formula for the joint distribution of 
formation masses and times.  
The structure of a halo is expected to be related to its accretion 
history.  Our tests show that our formulae for formation masses and 
times are reasonably accurate, so we expect that they will aid future 
analytic studies of halo structure.  
\end{abstract}


\section{Introduction}
There is a simple analytic approximation for the distribution of 
halo formation times, when formation is defined as the time when the 
most massive progenitor first contains at least half the mass of the 
final object (Lacey \& Cole 1993, 1994).  (Throughout, we will use the 
word parent to denote the final object, and the word progenitor to 
denote the smaller pieces which made up the mass of the parent 
at some earlier time.)  This formula provides a good description of 
what is seen in numerical simulations of gravitational clustering 
from Gaussian initial conditions, although recent work indicates that 
the agreement is not perfect (e.g., Wu 2001; Lin, Jing \& Lin 2003).  
The sense of the discrepancy is that haloes in simulations appear to 
form slightly earlier than predicted, in qualitative agreement with 
previous work by Tormen (1998).  

A related question is, what is the distribution of the mass of a halo 
at formation?  Absent other information, natural assumptions about 
this distribution are  
 (i) that it is a delta function centered at one-half, or 
 (ii) that the formation mass is uniformly distributed between one-half 
and unity.   
The second assumption is motivated by the fact that halo formation 
is expected to be a stochastic process; haloes of the same mass 
may have had different formation histories.  
The main purpose of the present paper is to derive and test a 
formula for the joint distribution of formation times and masses.  
Section~\ref{massform} studies the distribution of formation masses 
whatever the formation time.  It shows that the distribution of masses 
just prior to, and just after formation, measured in simulations are
both significantly different from delta functions, or from a uniform 
distribution, but are rather similar to simple formulae for these 
quantities derived by Nusser \& Sheth (1999).  
Section~\ref{joint} studies the conditional distribution of the 
formation mass, when the formation time is known.  This distribution 
is much better fit by a formula we derive here, than by a delta 
function or a uniform distribution.  
A final section summarizes our findings, and discusses 
possible applications.  

\section{The distribution of formation masses}\label{massform}
For what follows, it is useful to introduce some notation.  
We will use $\delta_{\rm sc}(z)$ to denote the value of the 
overdensity required for spherical collapse at $z$, extrapolated 
using linear theory to the present time (e.g. Peebles 1993), 
and $\sigma^2(m)$ will denote the variance in the initial density 
fluctuation field when smoothed with a tophat filter of comoving 
scale $R = (3m/4\pi\bar\rho)^{1/3}$, extrapolated using linear 
theory to the present time, where $\bar\rho$ is the comoving background 
density.  Thus, the shape of the initial power spectrum determines the 
relation between $\sigma$ and $m$.  
At any $z$, there is a characteristic mass scale defined by 
$\sigma^2(m)=\delta_{\rm sc}(z)$.  We will use $M_*(z)$ to denote this 
mass scale, and will often express masses in units of this 
characteristic mass.  

Later in this section we will compare our results with simulations; 
these were kindly made available to the public by the Virgo consortium 
(Frenk et al. 2000).  We will analyse results from the set of runs 
known as the GIF simulations.  In particular, we will show results from 
the SCDM and $\Lambda$CDM models, for which $\Lambda=1-\Omega$ and 
$(\Omega,\sigma_8)= (1,0.6)$ and $(0.3,0.9)$ respectively.  Particle 
positions and velocities from both simulations were output at a range 
of redshifts, approximately evenly spaced in logarithmic expansion 
factor: $\Delta\ln(1+z) \approx 0.0596$.  For each output time, we 
identified haloes using the spherical overdensity method 
(e.g. Lacey \& Cole 1994; Tormen, Moscardini \& Yoshida 2003)
which contained at least twenty particles.  The required overdensity 
is a cosmology dependent factor times the background density, as 
specified by the spherical collapse model.  For the SCDM model, this 
factor is 178, and it is independent of redshift; for the $\Lambda$CDM 
model, it is 323 at $z=0$, and is smaller at higher redshifts 
(e.g. Peebles 1993).  At any given output time $z_1$, we selected the 
halos which were composed of more than two hundred particles, and 
studied the formation times and masses at formation of these haloes as 
follows.  
(For reference, an $M_*$ halo at $z=0$, 0.5 and 1.0 has 
1289, 170 and 31 particles in the SCDM run, and 807, 185 and 40 
particles in the $\Lambda$CDM run, so the high redshift runs mainly 
probe the formation times and masses of objects much larger than $M_*$.)  

Given a halo of mass $M_1$ (i.e., containing $N_1$ particles) at $z_1$, 
we go to the previous output time ($z_1$+d$z_2$, say), identify the 
object which contributes the most number of particles to $N_1$, and 
call it the most massive progenitor at $z_1+{\rm d}z_2$.  Suppose this 
most massive progenitor had $N_2$ particles.  We then go to the 
preceding output step ($z_1+ {\rm d}z_2+{\rm d}z_3$, say) and identify 
the most massive progenitor, $N_3$, of $N_2$.  We continue in this way 
until the number of particles in the most massive progenitor first 
falls below $N_1/2$.  If the mass just before formation is $N_n$, then 
the mass just after formation is $N_{n-1}$, and the redshift of formation 
is $z_1+\cdots+ {\rm d}z_{n-1}$.  We store these values for each halo 
$M_1$ at $z_1$.  

The main quantity of interest in what follows is $p(m,z_{\rm f})$, 
the joint distribution of formation masses and times, where formation 
is defined to be the time when one of the subclumps of a halo first 
accounts for at least half the final mass $M_1$ of its parent.  
Because of this definition of formation, $m/M_1$ is distributed 
between one-half and unity (recall that the mass of the most 
massive progenitor must exceed half the mass of its parent).    

\begin{figure}
\centering
\epsfxsize=\hsize\epsffile{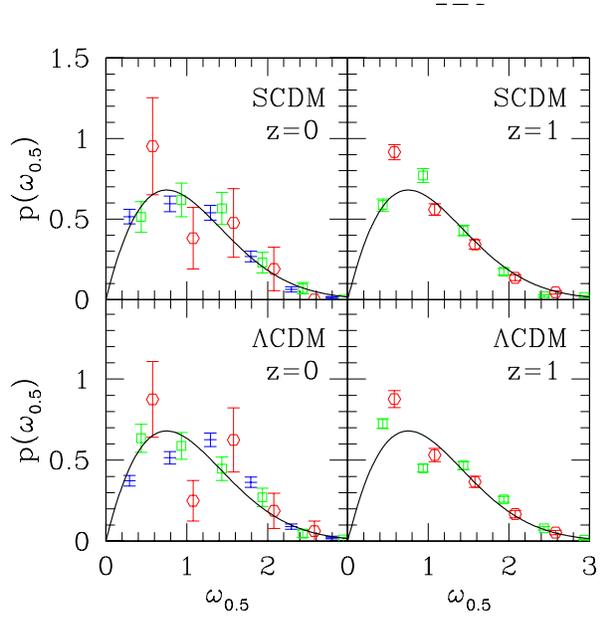}
\caption{Distribution of scaled formation times in two different 
 cosmological models, for haloes identified at two different redshifts.  
 In these scaled units, the formation time distribution is expected to 
 be independent of halo mass and final time.  
 Solid curve shows the precise form which this universal formation 
 time distribution is expected to have (equation~\ref{pwlc}).  
 In all panels, squares and hexagons show the simulation results for 
 parent haloes with masses in the range $4\le M_1/M_*(z_1)< 8$ and 
 $16\le M_1/M_*(z_1)< 32$.  Simple bars in the panels on the left  
 show results for slightly lower halo masses:  $M_1/M_*(z_1=0)\le 2$.  
 Error bars were estimated assuming Poisson counts.  
 Evidently, equation~(\ref{pwlc}) provides a reasonable, but not 
 perfect description of halo formation times in the simulations.  }
\label{ftimes}
\end{figure}

The formation time distribution of haloes which have final mass $M_1$ 
at redshift $z_1$, 
\begin{equation}
 p(z_{\rm f})\,{\rm d}z_{\rm f} = \int p(m,z_{\rm f})\,{\rm d}m,
 \label{pzdz}
\end{equation}
is expected to be well approximated by 
\begin{equation}
 p(z_{\rm f})\,{\rm d}z_{\rm f} = p(\omega)\,{\rm d}\omega 
  = 2\omega\,{\rm erfc}\left({\omega\over\sqrt{2}}\right)\,{\rm d}\omega 
 \label{pwlc}
\end{equation}
(Lacey \& Cole 1993), where 
$\omega^2 \equiv (\delta_{\rm cf}-\delta_{\rm c1})^2/(S_{\rm f}-S_1)$,
$\delta_{\rm cf}=\delta_{\rm sc}(z_{\rm f})$, 
$\delta_{\rm c1}=\delta_{\rm sc}(z_1)$, and 
$S_{\rm f}=\sigma^2(M_1/2)$.  As Lacey \& Cole note, this formula is 
only well-behaved for white-noise initial conditions (for which
$\sigma^2(m)\propto 1/m$), although it 
provides a reasonable approximation in the more general case.  

The distribution of formation masses, obtained by marginalizing over 
the formation time distribution, is 
\begin{equation}
 p(m)\,{\rm d}m = \int p(m,z_{\rm f})\,{\rm d}z_{\rm f}.  
 \label{pmdm}
\end{equation}
Nusser \& Sheth (1999) describe a model for the evolution of the 
mass of the most massive progenitor which is able to reproduce 
the formation time distribution of equation~(\ref{pwlc}).  
In their model, it is possible to derive an 
expression for the associated formation mass distribution 
of equation~(\ref{pmdm}).  In particular, they argue that 
\begin{equation}
 p(\mu)\,{\rm d}\mu = 
   {2\over\pi} \sqrt{1-\mu\over 2\mu-1}\,{{\rm d}\mu\over \mu^2},
   \qquad {\rm where}\ 1/2\le \mu\le 1,
 \label{pmformgt}
\end{equation}
and $\mu\equiv m/M_1$ (equation~A15 in Nusser \& Sheth 1999).  
Strictly speaking this formula, like equation~(\ref{pwlc}), is valid 
for white-noise initial conditions, but Nusser \& Sheth argued that 
it should provide a good approximation even if the initial spectrum 
has more large-scale power (see their Fig.~A2 and associated 
discussion).  

Figure~\ref{ftimes} compares the formation time formula, 
equation~(\ref{pwlc}), with measurements in the GIF simulations.  
(We have used the notation $\omega_{0.5}$ to emphasize that 
formation is when the largest progenitor subclump contains at 
least half the mass of the final parent halo.  Our requirement that 
parent haloes have at least two hundred particles means that we only 
probe the formation statistics of the most massive haloes at high 
redshift. )  
Although lower mass haloes identified at a given time tend to have 
formed at higher redshifts than more massive haloes 
(cf. Figure~\ref{zfmf} below), equation~(\ref{pwlc}) suggests that, 
when appropriately rescaled, all dependence on mass, time and the 
shape of the power spectrum should be removed.  The different panels 
in the figure show that the scaled formation time distributions in 
the SCDM and $\Lambda$CDM runs are reasonably, but not perfectly-well 
described by equation~(\ref{pwlc}).  

\begin{figure}
 \centering
 \epsfxsize=\hsize\epsffile{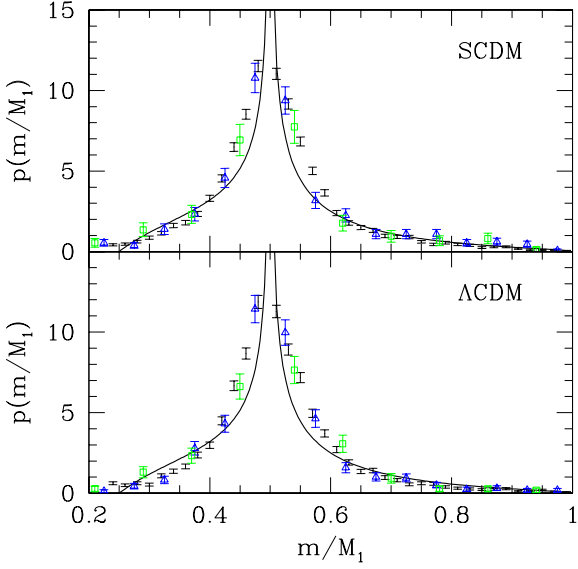}
 \caption{The distribution of masses $m$ at formation, for parent 
  haloes which have mass $M_1$ at $z_1=0$.  
  Symbols show the simulation results for $M_1/M_*(z_1)\le 1$ (dots), 
  $2\le M_1/M_*(z_1)< 4$ (triangles), and $M_1/M_*(z_1)\ge 8$ (squares).  
  Error bars were estimated assuming Poisson counts.  
  Curves on the right and the left of $m/M_1=1/2$ show the distributions 
  in equations~(\ref{pmformgt}) and~(\ref{pmformlt}) respectively.  
  There is no obvious trend with $M_1$, although haloes in simulations 
  appear to have $m/M_1\approx 1/2$ slightly more frequently than the 
  model predicts. Results for formation masses of parent haloes 
  identified at other redshifts are similar.  }
\label{fmass}
\end{figure}

Our next task is to test the accuracy of the formation mass formula, 
equation~(\ref{pmformgt}).  
The symbols in Figure~\ref{fmass} show the distribution of masses $m$ 
at formation for haloes in the GIF simulations which have final mass 
$M_1$ at $z_1=0$; dots show $M_1/M_*(z_1)< 1$, 
open triangles show $2\le M_1/M_*(z_1)< 4$, and squares show 
$M_1/M_*(z_1)> 8$.  
Error bars were estimated assuming Poisson counts.  
The figure shows no clear trend with $M_1$.  
A similar analysis of the formation masses, using haloes identified 
at $z=0.5$ and $z=1$, yields similar results.  
The solid curves which span the range $1/2\le m/M_1\le 1$ in the 
two panels of Fig.~\ref{fmass} show equation~(\ref{pmformgt}).  
Although the formation mass distribution measured in the simulations 
is significantly different from either a delta function, or a 
uniform distribution, equation~(\ref{pmformgt}) is able to provide 
a reasonable description of its shape.  

\begin{figure}
 \centering
 \epsfxsize=\hsize\epsffile{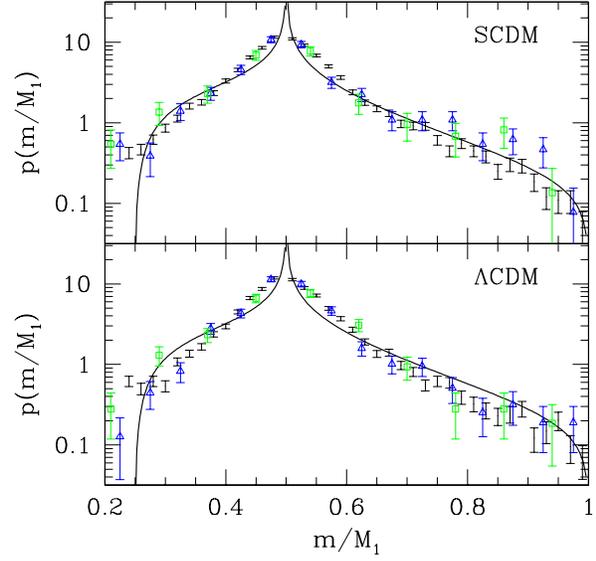}
 \caption{Same as previous plot, but now shown logarithmically, to 
  emphasize the discrepancy near the peak and in the tails.}
 \label{lgfmass}
\end{figure}

We also studied the mass of the most massive progenitor just before 
the formation time; these are shown by the symbols which span the 
range $1/4\le m/M_1\le 1/2$ in the two panels.  Once again, the 
measured distribution is neither a delta function nor is it 
uniform.  In this case, also, there is a simple formula for the 
distribution of formation masses:
\begin{equation}
 q(\mu)\,{\rm d}\mu = {{\rm d}\mu/\mu^2\over\pi(1-\mu)}
              \left(\sqrt{\mu\over 1-2\mu} - \sqrt{1 - 2\mu}\right),
 \label{pmformlt}
\end{equation}
where $1/4\le \mu\le 1/2$, and $\mu\equiv m/M_1$ as before 
(equation~A19 in Nusser \& Sheth 1999).
The curves which span the range $1/4\le m/M_1\le 1/2$ in the two panels 
of Fig.~\ref{fmass} show this formula; it provides a reasonable 
description of the measurements in the simulations.  

Although the analytic formulae provide a reasonable description of the 
measurements, haloes in the simulations appear to have slightly more 
occurences of $m/M_1\sim 0.45$, and $m/M_1\sim 0.55$ than the formulae 
predict.  Some of the discrepancy may arise because the simulation 
outputs are not spaced arbitrarily closely in time (the typical redshift 
steps are of order $\Delta z\sim 0.1$). As a result, the measured 
distributions almost certainly smooth-out the divergence around 
$\mu\sim 1/2$.  (To better illustrate the behaviour around the peak, 
Figure~\ref{lgfmass} shows the same distributions, but this time on a 
logarithmic scale.)  In principle, the analysis in Nusser \& Sheth (1999) 
can be used to estimate this smearing-out (their equations~A14 and~A18 
actually depend on the redshift difference), but we believe it will be 
better to use simulations with better time resolution instead, as these 
will be available shortly.  

Some of the discrepancy may be associated with the fact that the 
approach leads to an underestimate of the mean formation redshift.  
This discrepancy could plausibly affect the formation mass distribution, 
since, if formation happens at higher redshift when the basic building 
blocks are smaller, then the formation masses are less likely have 
values as large as $m/M_1\sim 1$.  
Moreover, equations~(\ref{pwlc})--(\ref{pmformlt}) are derived from an 
approach which predicts fewer massive parent haloes than are actually 
observed in simulations (e.g. Sheth \& Tormen 1999).  If the abundance 
of parent haloes is modified so that it is in better agreement with 
simulations, then the formation mass and time distributions will also 
be modified (for reasons made explicit in Sheth \& Tormen 2002).  
Accounting for this is left for future work, since the agreement between 
the model and the simulations is quite good.  

\section{Conditional distribution of formation mass and time}\label{joint}
The joint distribution of formation mass and time for parent haloes 
with mass $M_1$ at $z_1$ is 
\begin{equation}
 p(m,z)\,{\rm d}m\,{\rm d}z = {\rm d}s 
    \int_{S_{\rm f}}^{s_{\rm m}} \!\!\!\!{\rm d}S\,
                p(S,z+\Delta z|s,z)\,p(s,z|S_1,z_1),
 \label{pmzdmz}
\end{equation}
where $s\equiv\sigma^2(m)$, $S_1\equiv\sigma^2(M_1)$, 
$S_{\rm f}\equiv\sigma^2(M_1/2)$, $s_m\equiv\sigma^2(m/2)$, 
\begin{equation}
 p(s,z|S_0,z_0)\,{\rm d}s = {{\rm d}\nu\over\nu}\,
                            \sqrt{\nu\over 2\pi}\,\exp(-\nu/2),
\end{equation}
with $\nu\equiv [\delta_{\rm sc}(z)-\delta_{\rm sc}(z_1)]^2/(s-S_1)$, 
and a similar expression holds for $p(S,z+\Delta z|s,z)$.  
When inserted in equation~(\ref{pzdz}), equation~(\ref{pmzdmz}) yields 
equation~(\ref{pwlc}), and when inserted in equation~(\ref{pmdm}) it 
yields equation~(\ref{pmformgt}).

In the limit of small time steps ($\Delta z\ll 1$), and a white-noise 
power spectrum, equation~(\ref{pmzdmz}) simplifies considerably.  
A little algebra shows that, for haloes of fixed mass $M_1$, the 
conditional distribution of formation masses $m$ when it is known 
that the formation time was $z_{\rm f}$ is given by 
\begin{equation}
 p(\mu|z_{\rm f})\,{\rm d}\mu \equiv 
 {p(\mu,z_{\rm f})\,{\rm d}\mu \over p(z_{\rm f})}
  = {p(\mu)\,{\rm d}\mu\over s/S_1 - 1}\,
    {\exp\Bigl[-{\omega^2\over 2}{(S_{\rm f}-S_1)\over (s-S_1)}\Bigr]
     \over 2\,{\rm erfc}(\omega/\sqrt{2})},
 \label{pmz}
\end{equation}
where $\mu=m/M_1$, $s\equiv \sigma^2(m)$, $S_1\equiv \sigma^2(M_1)$, 
$S_{\rm f}\equiv\sigma^2(M_1/2)$, and $\omega$ was defined in 
equation~(\ref{pwlc}).  For a white-noise spectrum, $s/S_1 = 1/\mu$ 
and it is straightforward to verify that this distribution is 
correctly normalized.  For more general power spectra, 
$s/S_1 \sim \mu^{-\alpha}$, say, this conditional distribution 
must be multiplied by a normalization factor which depends on 
$\alpha$.  We have checked that use of the white-noise expression 
is a good approximation to the curves associated with $\alpha<1$, 
provided we insure that the distribution is correctly normalized to 
unity.  Thus, although equation~(\ref{pmz}) only holds for a 
white-noise power spectrum, we expect it to be more generally 
applicable for the same reasons that our equations~(\ref{pmformgt}) 
and~(\ref{pmformlt}) are more generally applicable.  In what follows, 
therefore, we will simply set $s/S_1 = 1/\mu$ and $S_{\rm f}/S_1=2$.  
In this approximation, our expression for the conditional distribution 
of formation masses is independent of power spectrum.  

The factor which multiplies $p(\mu)$ is largest at $s/S_1-1=\omega^2$, 
so objects which form at redshifts which are lower than the mean value 
for that mass (i.e., $\omega < 1$), are expected to have formation 
masses which are biased towards $\mu\approx 1$ (i.e., $s\approx S_1$).  
Conversely, objects which form at abnormally high redshifts 
($\omega>1$) are expected to have formation masses which are closer 
to the minimum value allowed:  $\mu\approx 1/2$.  Presumably, this 
is a consequence of the fact that, to have $\mu\approx 1$ requires 
two pieces each of size $\mu\approx 1/2$.  In a hierarchical model, 
the building blocks available to form the parent halo are, on average, 
smaller at early times:  when the probability of having an object of 
mass $\mu\approx 1/2$ is small, the chance of having two such objects 
is smaller still.  In effect, our formula~(\ref{pmz}) quantifies the 
importance of this effect.  

\begin{figure}
 \centering
 \epsfxsize=\hsize\epsffile{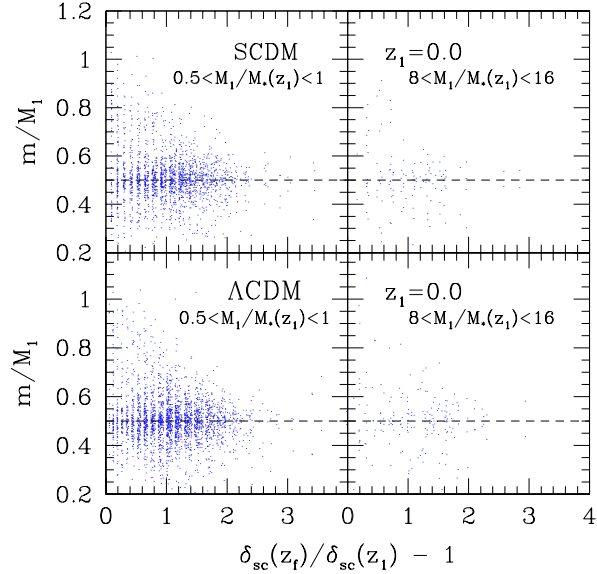}
 \caption{Joint distribution of formation times and masses measured 
  in the simulations.  Haloes which form at higher redshifts appear to 
  have a smaller spread in formation masses.  This is quantified in the 
  next figure, which includes a comparison with the model predictions.}
 \label{zfmf}
\end{figure}

Figure~\ref{zfmf} shows the joint distribution of formation mass and 
time for parent haloes identified at $z_1=0$ in the SCDM (top) and 
$\Lambda$CDM (bottom) simulations.  (The stripes are a result of the 
fact that simulation outputs are written to file only at finitely many 
time-steps.)  
The axis labels use the notation $m/M_1$ to denote the ratio of the 
formation mass to final mass, $z_{\rm f}$ the formation redshift, and 
$z_1$ the redshift at which the parent object was identified.  
The two panels for each simulation show results for different choices 
of the parent halo mass.  
Analogous plots for $z_1=0.5$ and $z_1=1.0$ look very similar, 
provided we scale the formation time axis to 
$\delta_{\rm sc}(z_{\rm f})/\delta_{\rm sc}(z_1) - 1$ as we have done, 
rather than simply show $z_{\rm f}$.  We have chosen to not include 
them here.  (The natural rescaling would have been to show $\omega$, 
defined in equation~\ref{pwlc}, along the x-axis.  This would differ 
from the rescaling we show by a factor of 
$(S_{\rm f}-S_1)/\delta_{\rm sc}(z_1)$.  
We chose not to scale by this additional factor because one of the 
points we wish to emphasize is that the formation mass formulae turn 
out to be approximately independent of power spectrum.)    

The formation time distribution discussed in the previous section is 
obtained by summing up all haloes with the same $z_{\rm f}$ whatever 
their value of $m/M_1 >1/2$.  The formation mass distributions studied 
in the previous section were obtained by summing up all haloes with the 
same $m/M_1$ whatever their value of $z_{\rm f}$.  

Notice that there appears to be a tendency for the objects with large 
$z_{\rm f}$ to have small values of $m/M_1$, but because there are many 
fewer haloes with high formation redshifts, it is not obvious if this 
trend is real, or if it is simply a consequence of small-number 
stastistics.  

\begin{figure}
 \centering
 \epsfxsize=\hsize\epsffile{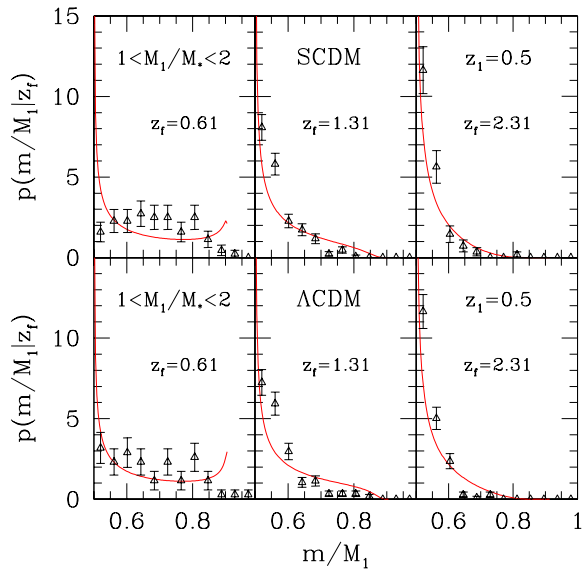}
 \caption{Conditional distribution of masses $m$ at formation, 
  given that the mass of the parent halo was in the range 
  $1<M_1/M_*(z_1)<2$ at $z_1=0.5$, for a range of choices of the redshift 
  of formation (labeled in the middle of each panel).  Symbols show the 
  measurements in the simulations, and curves show equation~(\ref{pmz}).}
 \label{mgivenz}
\end{figure}

To address this in more detail, Figure~\ref{mgivenz} shows 
$p(\mu|z_{\rm f})$, the conditional distribution of formation masses 
at fixed formation time.  The plot was made by choosing all haloes with 
masses in the range $1<M_1/M_*(z_1)<2$ at $z_1=0.5$, and then studying the 
mass at formation in the subset which formed at 
$z_{\rm f}=0.61$, 1.31 and 2.31.  
Histograms show the measurements in the simulations.  Comparison of the 
different panels shows that the objects which form at higher redshifts 
have formation masses which are close to $1/2$, whereas there is an 
obvious tail of higher formation masses at lower formation redshifts.  
The smooth curves show equation~(\ref{pmz}); it reproduces the trend 
with formation redshift seen in the simulations quite well.  
We find similar agreement for other choices of $M_1$, $z_1$ and 
$z_{\rm f}$, so we conclude that equation~(\ref{pmz}) provides a 
reasonable description (by which we mean it is a better fit than is 
a delta function, or a uniform distribution) of the conditional 
distribution of formation mass when the formation time is known.  

\section{Discussion}
We presented evidence that formulae for the distribution of formation 
masses (equations~\ref{pmformgt} and~\ref{pmformlt}), were reasonably 
accurate (Figure~\ref{fmass}).  
These formulae do not depend on the shape of the underlying power 
spectrum, so they are simple to use.  
We then derived an expression for the conditional distribution of 
formation masses if the formation time is known (equation~\ref{pmz}), 
and showed that it was also in quite good agreement with measurements 
made in simulations (Fig.~\ref{mgivenz}).  Application of Bayes' 
rule then gives the joint distribution of formation mass and time.  

Our results indicate that haloes which form at abnormally early times 
are more likely to have formation masses of order one-half that of 
the final mass of the parent, whereas haloes which form at abnormally 
late times are more likely to have formation masses which are closer 
to that of the parent.  One consequence of this is that haloes which 
form late are more likely to have experienced a recent major merger.  
We argued that this was a generic consequence of hierarchical formation.  

Our formulae for the joint distribution of formation masses and times 
will find use in studies which attempt to relate the structure of a 
halo to its formation history (e.g. Tormen 1997, 1998; 
Tormen, Diaferio \& Syer 1998; van den Bosch 2002; 
Wechsler et al. 2002; Zhao et al. 2003).  
For instance, haloes which formed recently with large formation masses 
are almost certainly further from equilibrium than haloes which formed 
at higher redshift with formation masses of order fifty-percent.  Such 
haloes (i.e. ones which have suffered major-mergers recently) may 
plausibly be less centrally concentrated than haloes of the same mass 
which had more quiescent accretion histories.  Addressing such issues 
is the subject of on-going work.  
If these formulae do prove to be useful, it will become necessary to 
modify them slightly so that they are more fully consistent with the 
parent halo mass function described by Sheth \& Tormen (1999).  

\bigskip

We would like to thank the Aspen Center for Physics for support, 
and for providing the stimulating environment in which this work 
was completed.  
We would also like to thank the Virgo consortium for making the 
simulation data used here publically available at   
{\tt http://www.mpa-garching.mpg.de/Virgo}.  
RKS was supported by the DOE and NASA grant NAG 5-7092 at Fermilab 
when work on this project began, and acknowledges support from  
NSF grant AST-0307747.

\end{document}